\documentstyle[epsfig,12pt,aasms4]{article}
\begin{document}

\lefthead{Fender et al.}
\righthead{GX339-4}
\slugcomment{Submitted to ApJ Letters}

\title{
Quenching of the radio jet during the X-ray high state of GX~339-4
}

\author{Robert Fender\altaffilmark{1}, St\'ephane Corbel\altaffilmark{2},
Tasso Tzioumis\altaffilmark{3}, Vince McIntyre\altaffilmark{4}, \\
Duncan Campbell-Wilson\altaffilmark{5}, Mike Nowak\altaffilmark{6},
Ravi Sood\altaffilmark{7}, Richard Hunstead\altaffilmark{4},
Alan Harmon\altaffilmark{8}, 
Philippe Durouchoux\altaffilmark{2}, 
William Heindl\altaffilmark{9} 
}

\altaffiltext{1}{Astronomical Institute
`Anton Pannekoek', University of Amsterdam, and Center for High Energy
Astrophysics, Kruislaan 403, 1098 SJ Amsterdam, The Netherlands}
\altaffiltext{2}{CEA-Saclay, Service d'Astrophysique, 91191 Gif sur
Yvette Cedex, France}
\altaffiltext{3}{Australia Telescope National Facility, CSIRO, Paul
Wild Observatory, Narrabri NSW 2390, Australia}
\altaffiltext{4}{Department of Astrophysics, School of Physics,
  University of Sydney, NSW 2006, Australia}
\altaffiltext{5}{Molonglo Radio Observatory, Department of Astrophysics,
School of Physics, University of Sydney, NSW 2006,  Australia}
\altaffiltext{6}{JILA, University of Colorado, Campus Box 440,
Boulder, CO 80309-0440}
\altaffiltext{7}{School of Physics, Australian Defence Force Academy, 
Canberra, ACT 2600, Australia}
\altaffiltext{8}{NASA/Marshall Space Flight Center, Huntsville, AL 35812}
\altaffiltext{9}{Center for Astrophysics and Space Sciences 0424, 
University of California at San Diego, La Jolla, CA 92093}

\begin{abstract}

We have observed the black hole candidate X-ray binary GX 339-4 at
radio wavelengths before, during and after the 1998 high/soft X-ray
state transition. We find that the radio emission from the system is
strongly correlated with the hard X-ray emission and is reduced by a
factor $\geq 25$ during the high/soft state compared to the more usual
low/hard state. At the points of state transition we note brief
periods of unusually optically-thin radio emission which may
correspond to discrete ejection events.  We propose that in the
low/hard state black hole X-ray binaries produce a quasi-continuous
outflow, in the high/soft state this outflow is suppressed, and that
state transitions often result in one or more discrete ejection
events. Future models for low/hard states, such as ADAF/ADIOS
solutions, need to take into account strong outflow of relativistic
electrons from the system.  We propose that the inferred Comptonising
corona and the base of the jet-like outflow are the same thing, based
upon the strong correlation between radio and hard X-ray emission in
GX 339-4 and other X-ray binaries, and the similarity in inferred
location and composition of these two components.

\end{abstract}


\section{Introduction}

GX 339-4 is one of only a handful of persistent black hole candidate
X-ray binaries known (Tanaka \& Lewin 1995). The system lies at a
distance of several kpc in direction of the Galactic centre
(e.g. Zdziarski et al. 1998) and exhibits a possible orbital
modulation with a 14.8 hr period in optical photometry (Callanan et
al. 1992), although this may in fact be half the true orbital period
(Soria, Wu \& Johnston 1999a).  GX 339-4 shares X-ray timing and
spectral properties with the classical black hole candidate Cyg X-1,
although exhibiting more frequent state changes and a larger dynamic
range of soft X-ray luminosity (Harmon et al. 1994; Tanaka \& Lewin
1995; M\'endez \& van der Klis 1997; Rubin et al. 1998; Zdziarski et
al. 1998; Nowak, Wilms \& Dove 1999; Wilms et al. 1999; Belloni et
al. 1999). The system is also a weak and persistent radio source with
flux densities typically in the range 5 -- 10 mJy at cm wavelengths
and a flat (spectral index $\alpha \sim 0$ where flux density $S_{\nu}
\propto \nu^{\alpha}$) spectrum (Fender et al. 1997; Corbel et
al. 1997; Hannikainen et al. 1998). The radio emission is roughly
correlated with both the soft (as observed with RXTE ASM) and hard (as
observed with CGRO BATSE) X-ray flux in the X-ray low/hard state
(Hannikainen et al. 1998). As discussed in Wilms et al. (1999) the
radio emission almost certainly arises in a region larger than the
binary separation, supporting an interpretation of its origin in a
compact partially self-absorbed jet, possibly of the type considered
by Hjellming \& Johnston (1988). Additional supporting evidence comes
from the recent resolution of a compact jet in VLBA observations of
Cyg X-1 (Stirling et al. 1998, de la Force et al. 1999), a source
whose radio, as well as X-ray, properties appear to parallel those of
GX 339-4 (Hannikainen et al. 1998; Pooley, Fender \& Brocksopp 1999)

\section{Observations}

\subsection{MOST}

Occasional monitoring of GX 339-4 wth the Molonglo Observatory
Synthesis Telescope (MOST) at 36 cm has been carried out for several
years.  All the observations were calibrated, imaged and CLEANed with
the standard MOST imaging pipeline (McIntyre \& Cram, 1999).  To
moderate any errors in the calibration we followed the procedure of
Hannikainen et al. (1998), fitting three sources besides GX339-4 in
each observation, and scaling the fluxes so that the sum of these
three reference sources remained constant, on the assumption that
these sources do not vary. The IMFIT task in the MIRIAD software
package (Sault, Teuben \& Wright 1995) was used to make point source
fits to the synthesised maps.  Further details, an observing log and
tabulated flux densities will be presented in Corbel et al. (1999);
see also Hannikainen et al. (1998). The MOST flux density measurements
are plotted in the top panels of Figs 1 and 2.

\subsection{ATCA}

Observations of GX 339-4 have been carried out at wavelengths of 21.7,
12.7, 6.2 and 3.5 cm with the Australia Telescope Compact Array
(ATCA).  Observational procedures are similar to those described in
Fender et al. (1997) and will be discussed more fully in Corbel et
al. (1999). Data reduction was performed with the MIRIAD software
package. The ATCA flux density measurements are plotted in the top
panels of Figs 1 and 2.

\subsection{CGRO BATSE}

The BATSE experiment aboard the Compton Gamma Ray Observatory monitors
the various hard X-ray sources in the sky using the Earth occultation
technique (Harmon et al. 1994). An optically thin thermal
bremsstrahlung model (with a fixed kT = 60 keV) has been used to fit
the data (following Rubin et al. 1998) and to produce the light curve
in the 20-100 keV energy band. We have checked for the presence of
bright interfering sources in the limb which could have biased the
measurement of the flux and have flagged suspicious data.  The 20-100
keV BATSE data are plotted in the middle panels of Figs 1 and 2.

\subsection{RXTE ASM}

GX 339-4 is monitored up to several times daily by the Rossi X-ray
Timing Experiment (RXTE) All-Sky Monitor (ASM) in the 2-12 keV
range. See e.g. Levine et al. (1996) for more details. The 2-12 keV
ASM data are plotted in the lower panels of Figs 1 and 2.

\section{Quenching of the radio emission}

Fig 1 plots the radio, hard- and soft-X-ray observations of GX 339-4
prior to, during and following the transition from the low/hard state
to the high/soft state in early 1998 January, and the transition back
to the low/hard state just over one year later (see Belloni et
al. 1999 for X-ray spectral and timing properties).  It is immediately
obvious that the radio and hard X-ray flux are strongly anticorrelated
with the soft X-rays and are consistent with zero measured flux for
the majority of the observations during the high/soft state. In
particular, there was no significant radio detection of GX 339-4
between MJD 50844 and the reappearance of the radio flux on MJD 51222,
despite eight observations with MOST at 843 MHz and three observations
with ATCA simultaneously at 4.8 and 8.6 GHz. The strongest limits on
the radio flux in the high state are the ATCA measurements which had a
typical $3\sigma$ flux density limit of $\sim 0.2$ mJy, constraining
the emitted flux density to be more than a factor of 25 weaker than
observed in the low/hard state. The single most stringent upper limit,
of 0.12 mJy ($3\sigma$), from the ATCA observation on MJD 51129,
constrained the radio flux to be more than forty times weaker than in
the low/hard state.

In Fig 2 we examine in more detail the period of state transition. In
addition we plot the low (1.3-3.0 keV) and high (5.0-12.2 keV) XTE ASM
channels, instead of simply the total intensity as in Fig 1; this
illustrates clearly the dramatic increase in the soft (disc) component
during the state transition.  Note that from XTE PCA timing
observations we can only be certain that by MJD 50828 the source was
in the high/soft state (Belloni et al. 1999). The most dramatic
decrease in the hard X-ray flux and corresponding increase in the soft
X-rays occur around MJD 50812 -- 50816 (centred on New Year
1997/1998).  By MJD 50822 the radio flux density had dropped to levels
undetectable with either MOST or ATCA ($3\sigma$ limit at 4.8 GHz of
$\leq 0.1$ mJy with ATCA) ; i.e. the timescale for decay from `normal'
to `quenched' levels is $\leq 10$ d. This is consistent with the
timescales for radio : X-ray correlations reported by Hannikainen et
al. (1998). However, subsequent radio observations revealed a small
resurgence in the radio flux density between MJD 50828 -- 50840 with
an unusually optically thin spectral index of $\sim -0.4$ (as measured
on MJD 50828). By MJD 50844 the radio flux had again dropped to
undetectable levels and was not detected again until over one year
later. The quenching of the radio emission simultaneously with a large
drop in the hard X-ray flux as observed with BATSE is reminiscent of
that observed in the radio-jet X-ray binary Cyg X-3 (McCollough et
al. 1999 and references therein).

\section{Reappearance of the radio emission}

Observations of GX 339-4 on MJD 51222 detected the radio source for
the first time in over a year (Fig 1).  The reappearance of the radio
source was coincident with the end of a long ($\geq 100$ d) decline
in the soft X-ray flux and a sharper increase in the hard X-ray
emission.  This return to the low/hard state was slow compared to the
corresponding transition by Cyg X-1 in 1996 which took $\leq 20$ d
(Zhang et al. 1997a).  As in the small pre-quenching flare event, the
spectral index immediately after the reapparance of the radio source
was unusually optically thin at around $-0.4$ (measured on MJD
51222). Subsequent observations have revealed a return to the flat
spectrum and steady flux densities previously observed in the low/hard
state. The timescale for the return from `quenched' to `normal' radio
states can only be constrained to be $\leq 20$ d.

\section{Discussion}

Our observations have revealed that the radio emission from GX 339-4
is strongly suppressed during the high/soft X-ray state. This
observation is in qualitative agreement with observations of an
increase in the strength of radio emission from Cyg X-1 during
transitions from the high/soft or intermediate states back to the more
common low/hard state (Tananbaum et al. 1972; Braes \& Miley 1976,
Zhang et al. 1997b). In addition, Corbel et al. (1999) present
evidence for previous periods of quenched radio emission in GX 339-4
which appear to correspond to periods of weak BATSE emission. We
assert that it is a characteristic of the high/soft state in black
hole X-ray binaries that radio emission is suppressed with respect to
the low/hard state. At least one model already exists for the
suppression of jet formation at high accretion rates in X-ray binaries
(Meier 1996), and may be relevant to this phenomenon.  It is unclear
at present how these findings relate to observations of radio emission
associated with X-ray transients in outburst (e.g. Hjellming \& Han
1995; Kuulkers et al. 1999) as (a) these sources may reach the
physically distinct Very High state (Miyamoto et al. 1991; Ebisawa et
al. 1994), and (b) the radio emission in these cases appears to
originate in discrete ejections, probably produced at points of X-ray
state change, and as such are decoupled from the system. We note that
Miyamoto \& Kitamoto (1991) have proposed a jet model for the Very
High state of GX 339-4.

In the low/hard X-ray state GX 339-4, in common with other black hole
candidates, does not display a strong soft (disc) component
(e.g. Wilms et al. 1999 and references therein). The inner regions of
the accretion flow may be described by an ADAF, ADIOS or `sphere +
disc' geometry (e.g. Narayan \& Yi 1995; Esin et al. 1998; Blandford
\& Begelman 1999; Wilms et al. 1999), all models in which the
standard, thin, accretion disc is truncated some distance from the
central black hole. A hot corona closer to the black hole Comptonises
soft photons to produce the observed hard X-ray emission.  In the
high/soft state the disc is believed to extend to within a few
gravitational radii of the black hole, resulting in a much increased
soft, thermal X-ray component with $kT \la 1$ keV. Simultaneously the
Comptonising corona is believed to shrink and cool, resulting in a
decrease and softening of the hard X-ray flux. Spectral fits to GX
339-4 data before and after the state transition under discussion are
in agreement with this scenario (Belloni et al. 1999).  In addition
Soria, Wu \& Johnston (1999b) present evidence that the outer
accretion disc/flow, responsible for optical emission lines, is
present in both the low/hard and high/soft states.

Adding our new observational constraint that the low/hard state
produces a radio-emitting outflow, and the high/soft state does not,
these models can be summarised qualitatively by a sketch such as Fig
3. The extremely strong correspondence between the hard X-rays and
radio emission in GX 339-4 and other X-ray binaries (e.g. GRO J1655-40
(sometimes), Harmon et al. 1995; GRS 1915+105, Harmon et al. 1997,
Fender et al. 1999; Cyg X-3, McCollough et al. 1999; Cyg X-1,
Brocksopp et al. 1999) suggests that the regions responsible for the
emission in the two energy regimes are strongly physically coupled. We
consider it likely therefore that the corona is simply the base of the
jet, and that the population of relativistic electrons responsible for
the radio emission (at some point further downstream in the outflow
when it becomes partially optically thin to cm radio emission) may be
the high-energy tail of the population of hot electrons responsible,
via Comptonisation, for the hard X-rays.

We note that it is possible that the outflow continues in the
high/soft state but that radio emission is not observed because of
greatly increased losses suffered by the relativistic electrons before
the flow becomes (partially) optically thin to radio emission.
In order for this to occur as a result of adiabatic expansion losses,
the ratio of lateral expansion rate to jet width would need to be $\ga
25$ times more in the high/soft state than in the low/hard state. In
order for synchrotron or inverse Compton losses to be responsible, an
increase by a factor of $\geq 25$ would be required in the magnetic 
($\propto B^2$) or radiation energy densities respectively. As current
models suggest adiabatic expansion is the dominant loss process in
conical jets (e.g. Hjellming \& Johnston 1988) the required increase
in the magnetic or radiation energy densities would probably need to
be even larger for these processes to result in quenching of the jet.

\section{Conclusions}

The radio emission from GX 339-4 is found to be strongly quenched in
the high/soft X-ray state, by a factor of $\geq 25$, in comparison to
the low/hard state. This quenching in radio emission is found to be
extremely well correlated with a decrease in hard ($\geq 20$ keV)
X-ray emission, suggesting a strong physical coupling between the
regions responsible for hard X-ray and radio emission. We propose that
high/soft states in black hole candidate X-ray binaries do not produce
radio-emitting outflows. Optically thin radio emission at the time of
transition to and from the high/soft state implies discrete ejections
of material at the point of state change, in agreement with
observations of X-ray transients and more unusual sources such as GRS
1915+105. However, many of those systems are observed in the Very High
or poorly-defined states, and the exact relation between their radio
emission and that of GX 339-4 is not well understood at present.  In
addition, the optically thin emission observed at these periods of
state transition is further evidence that the flat-spectrum radio
emission generally observed in the low/hard state results from
partially optically thick emission from a quasi-continuous jet.  The
dramatic coupling between the emission from the inner ($\leq$ few 100
km) accretion disc and the radio emission is further confirmation that
jets are generated close to the compact object.

Physical models developed to interpret the low/hard states (ADAF,
ADIOS, sphere + disc) clearly need to take into account the direct
evidence for a continuous outflow in these states. Models for jet
formation need to consider why such accretion geometries produce
outflows whereas those envisaged to explain the high/soft state do
not, and why discrete ejection events are often, perhaps always,
observed at the point of state transitions in X-ray binaries.

\acknowledgements

RPF thanks Mariano M\'endez, Eric Ford, Tomaso Belloni and Michiel van
der Klis for useful discussions. The Australia Telescope is funded by
the Commonwealth of Australia for operation as a National Facility
managed by CSIRO.  MOST is operated by the University of Sydney and
funded by grants from the Australian Research Council.  RXTE ASM
results provided by the ASM/RXTE teams at MIT and at the RXTE SOF and
GOF at NASA's GSFC.  RPF was funded during the period of this research
by ASTRON grant 781-76-017 and EC Marie Curie Fellowship ERBFMBICT
972436. MN was supported in part by NASA Grant NAG5-3225 and NSF Grant
PHY94-07194.

\clearpage

\figcaption{Radio, hard- and soft-X-ray monitoring of GX 339-4 before,
during and after the 1998 high/soft state. The radio emission is
reduced by more than a factor of 25 during this state. Unsually
optically thin emission is observed just before and after the state
changes, probably corresponding to discrete ejection events.}

\figcaption{
As Fig 1, but with the timescale around the state transition expanded
and the lowest and highest energy XTE ASM channels plotted to
illustrate the abrupt rise in the soft (disc) component.}

\figcaption{
A sketch qualitatively combining the observational data with favoured
current models for accretion modes. In the low state (when the inner
accretion region may be advection-dominated) there is a
quasi-continuous radio jet, in the high state (when the hot thin
accretion disc extends much closer to the black hole) the radio jet is
not present.}

\begin{figure*}
\centerline{
\epsfig{file=highstate-bw.ps,angle=270,width=12cm,clip}
}
\end{figure*}

\clearpage

\begin{figure}
\centerline{
\epsfig{file=low2high.ps,angle=0,width=10cm,clip}
}
\end{figure}

\clearpage

\begin{figure*}
\centerline{
\epsfig{file=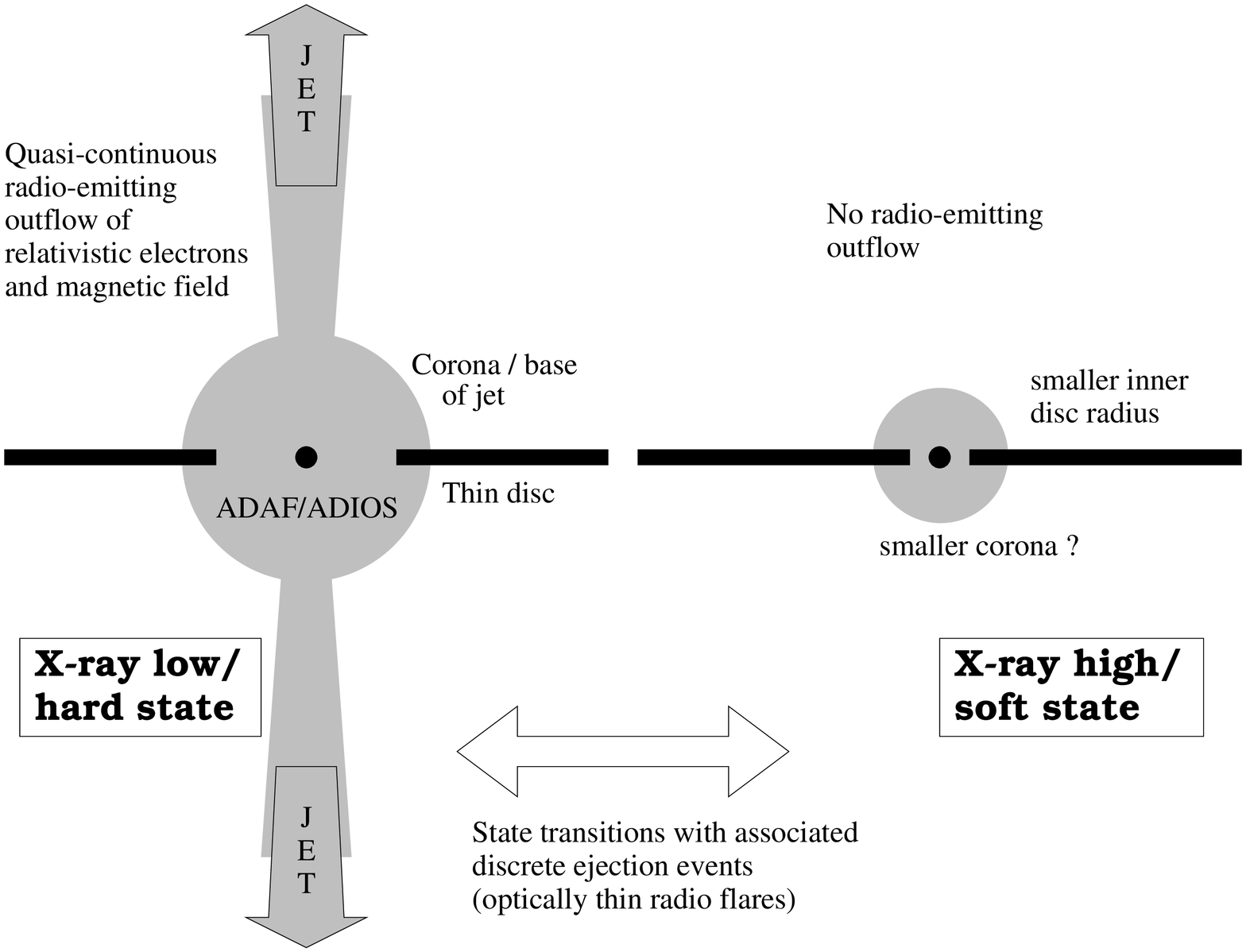,angle=270,width=12cm,clip}
}
\end{figure*}

\end{document}